\documentstyle[12pt]{article}
\newcommand{\be}{\begin{equation}} \newcommand{\ee}{\end{equation}}
\newcommand{\bea}{\begin{eqnarray}}\newcommand{\eea}{\end{eqnarray}}

\begin{document}
\begin{titlepage}

\title { Enhancement of Superconducting correlation due to
interlayer tunneling}
\author{Haranath Ghosh$^1$ and M. Sardar.$^2$ \\
%footnotemark[2] \\
Institute of Physics, Sachivalaya Marg, \\
Bhubaneswar-751005, INDIA.}
\footnotetext[1]{e-mail: \ hng@iopb.ernet.in}
\footnotetext[2]{e-mail: \ manas@iopb.ernet.in}
\maketitle
\thispagestyle{empty}

\begin{abstract}
Interlayer single particle tunneling between the $Cu-O$ layers suppress the
in-plane short range magnetic order (which is modeled as spin density
wave (SDW) insulator). Doping over the SDW state kills
perfect nesting of the Fermi surface (FS) in certain directions and hence SDW
gap reduces to zero in those directions of the FS. Coupling
between the planes through
interlayer tunneling ($t_{\perp}$) further suppresses the
in-plane magnetic
SDW-gap and hence becomes anisotropic.
Superconductivity arises in the gapless regions of the FS under
the `modified spin bag' mechanism. We show that the highest
$T_c$ can only be obtained for non-zero $t_{\perp}$ based on
this mechanism.

\end{abstract}

\vspace{.1in}
\hspace{0.12in}{\bf Keywords :} Spin density wave, Pseudo gap,
Anisotropic superconductors, Pair breaking.

\end{titlepage}
\eject

\section{Introduction.}
Short range antiferromagnetic correlations in the
superconducting (SC) phase of high $T_c$ materials have been
confirmed by numerous
neutron \cite{1} and Raman Scattering experiments \cite{2}, although the proper
understanding of the nature of the SC mechanism is far from reality.
However, the normal state of the cuprates at half filling is
insulating and antiferromagnetic in nature. There are two different
ways of describing the insulating state of the pure (undoped)
material : (a) the localised description of the electrons in the
Mott--Hubbard insulator for which $ U >> W$ (bandwidth), the
system stabilises to the so called Resonating Valence Bond ( RVB
)\cite{3} state in comparison to the Neel ordered ground state.
(b) In the limit $U~\le~W$, i.e in the itenarent description of the
electrons  the nesting of the FS leads to the SDW or CDW instability
driving the system from normal metallic to insulating state \cite{sch}.
We explore the later possibility. The
mechanism of SC in the two descriptions are expected to be of
different origin, as the nature of charge carriers and the effective
pairing interactions are quite different.

The important features of band structure calculations for 214 and 123
compounds are the following \cite{4} ;
$(i)$ The bands show very little dispersion along the c-axis,
indicating the 2-d nature of the electronic states,  $(ii)$ The Fermi
surface is  square planar with perfect nesting along the wave vector
$\vec Q=({\pi\over a},{\pi \over a}, 0)$. The later fact further confirms
the low dimensional character of electronic states and predicts the
instability of the system against the formation of a spin density
wave (SDW) state. Based on such a physical picture Schrieffer et al,
\cite{5} describes the ground state of the cuprates as SDW insulator,
characterised by an isotropic SDW gap through out the Fermi surface
(FS). Doping with injected charge carriers essentially suppress the SDW
gap locally which acts as a potential well for the carriers, in which
the injected charge carrier gets self trapped (so called spin
bag). Two such holes of opposite
spins find it energetically favourable to stay together digging a deeper
well. This is so, because there always exists an attractive
interaction between the charge carriers mediated by the quanta of
fluctuations of the amplitude mode of the SDW gap\cite{5,6} and hence
superconductivity sets in.

However, superconductivity under such a situation is restricted
either to lower or
the upper SDW band.
%But the predicted gap mode in this model is not
%seen in the Raman spectra\cite{6}.
Such a picture is physically
reasonable strictly in very low doping limit, so that the nesting of
the FS remains intact. On the other hand
as one dopes over the SDW ground state, it is expected that the
FS will move away
from perfect nesting and thereby the the SDW gap will no longer be an
isotropic one. The SDW gap will exist only in the directions of the
FS, where nesting is still preserved and with no gap in other directions,
resulting in a pseudo gap rather than a real gap at the Fermi level.
Therefore, the above picture needs necessary modifications as will be
described in more details in the next section.

Experimentally \cite{1} it is found that both the short range spin ordering as
well as the dispersion of the spin excitations above the Neel
temperature is predominantly 2-d in nature. This is so, because the
exchange integral between the planes ( proportional to the square of
the interlayer tunneling matrix element $t_{\perp}$) is much smaller
$\sim$ $10^{-5}$ compared to the inplane exchange coupling.
On the other hand, in our `modified spin bag' scenario, there are low
energy quasiparticles available in large parts of the FS, where
nesting is lost. They are now available to tunnel across the planes.
Hence the interplane tunneling of quasiparticles (linear in
$t_{\perp}$) is expected to be a more relevant perturbation than the
interplane exchange coupling.

  In the present paper, our main aim is to incorporate a physically
realistic model for high $T_c$ superconductivity modifying
Schrieffer's spin bag model\cite{5}, which holds good only in the
low doping limit. Subsequently, emphasis has been given to interlayer tunneling
of SDW quasiparticles in the intermediate metallic SDW state. We find
that the interlayer tunneling between the planes, suppress the
inplane SDW gap, which in turn enhances the superconducting
correlations.
In sec.II, we describe the basic ingradients of the
`modified spin bag', indicating the shortcomings of the
spin bag model. Sec. III is devoted
to show, how the interlayer tunneling between the layers
suppress the inplane magnetic SDW gap. The modified spin bag model is employed
to describe superconductivity in sec. IV. Finally, we remark about our results
in the concluding section V.
\section{Modified Spin Bag Model.}
The normal state of the high $T_c$ superconductors exhibits a number
of anomalous properties like, linear temperature dependence of
resistivity, Raman continuum \cite{7} due to scattering
of light by charge carrier. In order to
understand such puzzling features of the cuprates in the normal as
well as the SC state, realistic theoretical model is necessary. One
of the most important theoretical model in this direction is the spin
bag model of Schrieffer et al\cite{5} as mentioned in the last
section. According to spin bag model, on doping holes the long range
Neel order is supressed, but still short range magnetic order
persists for small doping away from half filling. This can be
identified with a SDW state (however in dispute). The Cu-O
planes being 2-d square planar
exhibits perfect nesting \cite{8}. Presence of high degree of nesting
 in the Fermi surface, together with strong in-plane spin correlation
between the $Cu$-spins drive
the system to SDW instability and in the process there is a net gain
in energy, because states with higher energies come down to lower
energy states resulting in a gap at the Fermi level of the electronic
spectrum. Now, in order to get SC one needs to dope the system with
charge carriers. In very low doping limit, doping supresses the SDW
gap locally and hence Schrieffer's spin bag picture holds
good. However, for all practical purposes, superconductivity in these
materials appear only at larger doping over the Neel ordered state. Hence
the FS is expected to deviate from perfect nesting. As a result the
SDW gap will vanish in certain directions, rather than just the local
supression as in the spin bag model. Therefore, the SDW gap will
exist only in certain parts of FS in which nesting still exists,
where as in other directions there is no gap, resulting in a pseudo
gap for excitations. Under these circumstances the pairing takes place
dominantly between the SDW quasiparticles from both the bands
(lower and upper SDW bands). The pairing interaction between the
quasiparticles in the regions of the FS, where the gap is
supressed is mediated by the quanta of the
fluctuations of the amplitude and phase modes of the SDW gap (
collective modes ). The presence of SDW gap in certain parts of the
FS is however essential, because it is the quanta of the fluctuations of the
collective modes that mediate the pairing interaction between
the quasiparticles.
The main point being that here the superconductivity is not
restricted either to the SDW valence or conduction band alone, but is a
global phenomena, unlike in \cite{5}.
The low energy SDW quasiparticles, that feel the pairing
interaction, come from the regions where SDW gap has gone to zero
or vanishingly small, i. e, from the regions where the lower (valence)
and upper (conduction) SDW bands touch other. So they come from both
the valence
and conduction bands.
These are the basic
ingredients of the ``modified spin bag model " ( a similar model
for non cuprate systems called the
``modified charge bag model" is proposed in \cite{hng}).

However, this is a purely two dimensional picture where the coupling
between the planes is neglected. On the other hand, under the modified spin
bag scenario, the nature of the SDW quasi particles (above $T_c$) in
most of the regions of the FS being free electron like, the interlayer
tunneling between planes is likely to play a crucial role.
The motivation for introducing the
coupling between the planes through interlayer tunneling is as
follows. Anisotropy in the normal state properties (resistivity,
thermal conductivity) is typically of the order of $10^3$ or more,
whereas typical anisotropy of quantities like coherence length,
penetration depth below  $T_c$ is $\approx 5-10$. Also the
analysis of the fluctuation conductivity and diamagnetic fluctuations
above  $T_c$ has shown that the character of fluctuation changes
from 2-d to 3-d as one approaches  $T_c$. This shows that the
normal to SC transition is at the same time a dimensional crossover
from 2-d to 3-d. This is taken with the fact that the BISCO materials
($T_c~ \sim 10^o K$  for single layer, $\sim 80^o $ K for double
layer and $\sim 110^o $ K for
triple layers ) clearly indicates that the third
direction coupling between the planes,
is a relevant parameter that ultimately brings in  the
full 3-d coherence. The $T_c$ may not be a single plane property and
interplanar coupling has to be ultimately incorporated in any kind of
theory of superconductivity \cite{and}. Here we explore the effect of
single particle tunneling on the metallic SDW ground state (the normal
state) of the $Cu-O$ planes and study the observable effects on
the SC state.

Usually in coupled planar superconductors, the tunneling of single
particles between the planes, suppress superconductivity, because as far as
individual planes are concerned, it acts as a pair breaking
perturbation, and hence $T_c$ is reduced. In contrast, here we
will show that, under this novel mechanism of superconductivity
exactly the opposite happens and maximum $T_c$ is got only for non zero
$t_{\perp}$
between the planes.

\section{Effect of Interlayer tunneling on the SDW state.}
A mean field theory for the SDW state starting
from the original Hubbard interaction has already been well
studied \cite{5}. Perfect nesting of the FS exists due to the
2-d nature of the electrons in the Cu-O plane, and is described by the
tight binding dispersion relation,
\bea
\epsilon_k=-\epsilon_{k+Q} =-2 t ( {\rm cos}k_xa +{\rm cos}k_ya)
\eea
where $Q$ being the nesting wave vector corresponding to the 2-d
square lattice.
In contrast, (to \cite{5,6}) we consider two Cu-O layers per unit
cell, where each $Cu-O$ layer can be described as SDW insulator in the ground
state. Doping over the SDW state destroys FS nesting in certain directions,
resulting in the suppression of the SDW gap. The behaviour of quasi particles
in the regions where nesting is lost, is likely to be a nested Fermi liquid
\cite{ruv}. Therefore, the interlayer tunneling between the metallic $Cu-O$
layers is likely to
play an important role in the subsequent SDW and hence SC state. We model our
system as two SDW planes
corresponding to two layers, coupled by single particle
interlayer tunneling.

$$H  = H_o + H^{\prime}$$ \\
\be
H_0  =  \sum_{k\sigma}
(\epsilon_k-\mu)(c_{1k\sigma}^{\dagger}c_{1k\sigma}
-c_{1k+Q\sigma}^{\dagger}c_{1k+Q\sigma}
) +  G \sum_{k\sigma} (\sigma_3)_{\sigma\sigma^{\prime}}
(c_{1k+Q\sigma}^{\dagger} c_{1k\sigma} + {\rm h.c })
  +  1\rightarrow 2
\ee
\be
H^{\prime}=\sum_{k\sigma}t_{\perp}(c_{1k\sigma}^{\dagger}c_{1k\sigma}
+ {\rm h.c} )
\ee

where the indices 1 and 2 refers to two different layers, $t_{\perp}$
and $ G$ being the tunneling matrix element and the SDW order
parameter respectively. The Hamiltonian $H_0$ in (2) describes the SDW states
of different layers obtained by the mean field calculation of the Hubbard
model \cite{5,6}. The interlayer tunneling between the layers, treated as
small perturbation over the unperturbed SDW state ($H_0$), is described
in equation (3). The SDW order parameter in different
layers are defined as,
\begin{eqnarray}
 G  =  -{1\over 2} {U\over N} \sum_{k\sigma}^{FBZ} \langle
c_{1k+Q\sigma}^{\dagger} (\sigma_3)_{\sigma\sigma^{\prime}}
c_{1k\sigma}\rangle =-{1\over2}{U\over N} \sum_{k\sigma}^{FBZ}
\langle
c_{2k+Q\sigma}^{\dagger}(\sigma_3)_{\sigma\sigma^{\prime}}
c_{2k\sigma} \rangle
\end{eqnarray}

where $U$ being the strength of the onsite Coulomb repulsion. Equation (4)
indicates that each layer has the same SDW gap. The $k$-sum is over the full
Brillouin zone. On the other hand the $k$-sums in equations (2), (3) are
extended only upto the reduced magnetic Brillouin zone bondary.

Usually, for quasi 1-d systems the transition to SDW (repulsive
system of electrons) or CDW (with strong electron phonon
interaction) are calculated approximately within the framework
of purely 1-d models, for which the exact $T_c$ should ,
according to the Landau theory be exactly zero. To get a nonzero
$T_c$ one needs to introduce transverse coupling between the
chains or planes, which could be either of two kinds, coulomb
reulsion between the chains, or kinetic coupling, i.e single
particle hopping ( tunneling ) term between the chains or planes.
The presence of non zero $t_{\perp}$ in the third direction usually
stabilizes long range order at finite temperatures
 for quasi 1-d or quasi 2-d systems.
So sufficiently close to the critical point, the 3-d character of
 correlation starts developing, so that a Ginzberg Landau
theory of phase transition becomes valid.
Two quite incompatible approaches are used to deal with anisotropic
systems. One being, the existence of finite $t_{\perp}$
and $T_c$ justifies the use of global mean field, that neglects
all lower dimensional effects.
In the other scheme, one does mean field in the transverse
direction only, treating the lower dimensional system more
rigorously. The later approach has the merit of correctly
reproducing the relevant physics in the limit $t_{\perp} \rightarrow
0$. However , it has very restricted domain of application\cite{fir}.
Reason is that, the transverse mean field is usually
done on an effective interplane two particle hopping term, whereas
the bare Hamiltonian has only single particle tunneling
process. Standard perturbation theory in $t_{\perp}$
will lead to an effective pair hopping across the planes with
amplitude proportional to $t_{\perp}^2/ \Delta$, only if the
two particles or a particle and hole are bound with a gap
equal to $\Delta$, in the lower dimensional spectrum.
In this way , there is no real one particle hopping, and it
is through virtual processes that the transverse pair motion
is possible.
On the other hand, in our mechanism as we shall discuss
later on, the SDW gap is zero
over some parts of the Fermi surface and hence the gap becomes
irrelevant. The ratio $t_{\perp}/\Delta$ is not small any more
and perturbation theory breaks down.
The  effect of small $t_{\perp}$ in the pseudo gap situation
has to be handled more carefully. Here we include and treat
the single particle hopping exactly, and ignore the possible
 generation of particle hole pair hopping across the planes.

We rewrite the Hamiltinian (2) in the matrix representation as folows,
\be
H=\sum_{k \sigma} \Psi_{k\sigma}^{\dagger} \hat H_{M} \Psi_{k\sigma}
\ee
where the Hamiltonian matrix ( $ \hat H_{M}$ ) and four component Nambu
operators are obtained as,
\bea
\hat H_{M}=\pmatrix {
(\epsilon_k-\mu)  & G(\sigma_3)_{\sigma\sigma}  &t_{\perp}  &0\cr
G (\sigma_3)_{\sigma\sigma}  &-(\epsilon_k-\mu)  & 0 &0 \cr
t_{\perp}  &0  &(\epsilon_k-\mu)  & G(\sigma_3)_{\sigma\sigma}\cr
0  &0  & G(\sigma_3)_{\sigma\sigma}  &-(\epsilon_k-\mu)
\cr }
\eea

and
\be
\Psi_{k\sigma}^{\dagger} = ( c_{1k\sigma}^{\dagger} ~~~c_{1 k+Q\sigma}^{
\dagger} ~~~c_{2k\sigma}^{\dagger} ~~~c_{2 k+Q\sigma}^{\dagger} )
\ee

In order to diagonalize the Hamiltonian (2) one needs to find a suitable
Bogoliubov transformation. To do so, we calculate the eigenvalues and
eigenvectors of the matrix $ \hat H_{M}$ by unitary transformation, such that,
$ H_{diag} = \hat U_{k\sigma}^{-1} \hat H_{M} \hat U_{k\sigma} $
and hence the diagonalised SDW Hamiltonian can be obtained as,
\be
H= \sum_{k\sigma} \phi_{k\sigma}^{\dagger} H_{diag} \phi _{k\sigma}
\ee
where $ \phi_{k\sigma}^{\dagger} =( \gamma_{1k\sigma}^{c\dagger}, \gamma_{
1k\sigma}^{v\dagger},\gamma_{2k\sigma}^{c\dagger},\gamma_{2k\sigma}^{
v\dagger})$,
are the new SDW basis states, that diagonalise the Hamiltonian
(2).$\gamma_{1,2}^{c,v}$ are the SDW quasiparticle annihilation
operators in the
conduction (valence) band for the effective nonbonding ( bonding )
combinations(1,2).

	Now, in order to find out the explicit structure of the
Hamiltonian (8) we need to find out explicit structure of $\hat
U_{k\sigma}~(\Psi_{k,\sigma}=\hat U_{k\sigma}\phi_{k\sigma})$ which
requires the evaluation of the eigen vectors of
$\hat H_M$ because each column of $\hat U_{k,\sigma}$ is given
by the respective ortho-normalised eigen vectors of $\hat
H_{M}$. We evaluate the unnormalised orthogonal eigen vectors of
 $\hat H_M$ as,
\bea
\hat e_1 & = &\pmatrix{1 \cr (E_{k}^0 - \epsilon_{k}^0)/\hat
G(\sigma^3)_{\sigma \sigma^\prime} \cr -1 \cr -(E_{k}^0 -
\epsilon_{k}^0)/\hat
G(\sigma^3)_{\sigma \sigma^\prime} \cr},
\hat e_2=\pmatrix{1 \cr -(E_{k}^0 + \epsilon_{k}^0)/\hat
G(\sigma^3)_{\sigma \sigma^\prime} \cr -1 \cr (E_{k}^0 +
\epsilon_{k}^0)/\hat
G(\sigma^3)_{\sigma \sigma^\prime} \cr},
\hat e_3=\pmatrix{1 \cr -(\tilde E_{k}^0 + \tilde\epsilon_{k}^0)/\hat
G(\sigma^3)_{\sigma \sigma^\prime} \cr 1 \cr -(\tilde E_{k}^0 +
\tilde\epsilon_{k}^0)/\hat
G(\sigma^3)_{\sigma \sigma^\prime} \cr}, \nonumber \\
& &
\hat e_4=\pmatrix{1 \cr (\tilde E_{k}^0 - \tilde\epsilon_{k}^0)/\hat
G(\sigma^3)_{\sigma \sigma^\prime} \cr 1 \cr (\tilde E_{k}^0 -
\tilde\epsilon_{k}^0)/\hat
G(\sigma^3)_{\sigma \sigma^\prime} \cr}
\eea

	Therefore, one can readily find the transformation
matrix $\hat U_{k,\sigma}$ (or the so called appropriate
Bogoliubov transformation) that connects the 4-component
Nambu-operator $\Psi_{k,\sigma}$ of the original lattice to the
4-component Nambu operator $\phi_{k,\sigma}$ for the condensate
SDW state as,
\bea
\pmatrix{c_{1 k \sigma} \cr c_{1 k+Q \sigma} \cr c_{2 k \sigma}
\cr c_{2 k+Q \sigma} \cr} = \pmatrix{u_{k}^1 & v_{k}^1 & v_{k}^2
& u_{k}^2 \cr (\sigma^3)_{\sigma\sigma^\prime}v_{k}^1 &
-(\sigma^3)_{\sigma\sigma^\prime}u_{k}^1 &
-(\sigma^3)_{\sigma\sigma^\prime}u_{k}^2 &
(\sigma^3)_{\sigma\sigma^\prime}v_{k}^2 \cr - u_{k}^1 &
-v_{k}^1 &  v_{k}^2 &  u_{k}^2 \cr -
(\sigma^3)_{\sigma\sigma^\prime}v_{k}^1 &
(\sigma^3)_{\sigma\sigma^\prime}u_{k}^1 & -
(\sigma^3)_{\sigma\sigma^\prime}u_{k}^2 &
(\sigma^3)_{\sigma\sigma^\prime}v_{k}^2 \cr}
\pmatrix{\gamma_{1k\sigma}^c \cr \gamma_{1k\sigma}^v \cr
\gamma_{2k\sigma}^v \cr \gamma_{2k\sigma}^v \cr}
\eea

Where $u_k^1 (v_k^1) ={1\over2} (1\pm {\epsilon_k^0\over E_k^0})^{1/2}$
and $u_k^2 (v_k^2) ={1\over2} (1\pm {\tilde\epsilon_k^0\over
{\tilde E_k^0}})^{1/2}$
, with $E_k^0$ and $\tilde E_k^0$ being the SDW quasiparticle energies for the
antibonding and bonding bands, and are given by,
\be
E_k^0 = \sqrt {(\epsilon_k^0)^2 +  G^2 }=\sqrt
{(\epsilon_k-\mu -{t_{\perp} \over 2})^2 +  G^2 }
\ee
\be
\tilde {E_k^0} = \sqrt {\tilde\epsilon_k^{0^2} +  G^2 }=\sqrt {(\epsilon_k-\mu
+{t_{\perp}
\over 2})^2 +  G^2 }
\ee

Hence the mean field SDW Hamiltonian for the system can be obtained using
equations (5 - 10) as ,
\be
H_{SDW}  = \sum_{i=1}^{2} \sum_{k\sigma}(E_{ik}^c  \gamma_{ik\sigma}^{c
\dagger}\gamma_{ik\sigma}^c +
E_{ik}^v\gamma_{ik\sigma}^{v\dagger}\gamma_{ik\sigma}
^v )
\ee

 where i=1,2 , corresponds to the bonding and antibonding SDW bands and
\bea
E_{1k}^{c(v)}={-t_{\perp}\over 2} \pm E_k^0
{}~~~ {\rm and} ~~~
E_{2k}^{c(v)}={t_{\perp}\over 2} \pm \tilde {E_k^0}
\eea
are the quasi particle energies in the respective bands.
Note that $E_{1k}^{c(v)}$ and $E_{2k}^{c(v)}$ are symmetric about
$-{t_{\perp}\over 2}$
and ${t_{\perp}\over 2}$ respectively.
In absence of $t_{\perp}$,
both the layers are degenerate with SDW quasiparticle
energies $ E_k=\pm \sqrt {(\epsilon_k-\mu)^2 + G^2 }$ and
both the SDW planes are equally gapped.
On the other hand, for any finite $t_{\perp}$ the hybridized
effective bands are non degenerate, being symmetric only about
the shifted Fermi energy.
This clearly indicates that the interlayer tunneling have non trivial
observable effects on the SDW ground state. The remaining part
of this section will
be devoted to showing how the SDW gap gets modified in
presence of interlayer tunneling.

Now the actual nature of the SDW gap parameter can be found out from the self
consistent calculation of the SDW order parameter given in eqn (4) using
eqn (10).
\be
 G = -{U\over N}{1\over2} \sum_{k\sigma}\sum_{i=1}^{2} (u_k^i v_k^i) ( \langle
\gamma_{ik\sigma}^{c\dagger}\gamma_{ik\sigma}^c \rangle - \langle
\gamma_{ik\sigma}^{v\dagger} \gamma_{ik\sigma}^v \rangle )
\ee

Where the thermal averages $<...>$ have to be evaluated with the mean field
Hamiltonian (13 ). The resulting gap equation can be written as,
\be
1= -{U\over 2} \sum_k {1\over 2E_k^0 } [f(E_{k}^1) - f(E_{k}^2)]
 -{U\over 2} \sum_k {1\over{2\tilde E_{k}^0}}[f(E_{k}^3) -
f(E_{k}^4)]
\ee

For  T=0 the above gap equation (16) simplifies to,
\be
1= -{U\over 4} \sum_k ( {1\over E_k^0} + {1\over \tilde E_k^{0}} )
\ee
Exact solution of the gap equation (17) results,
\be
 G ( T=0 ) = 2 \sqrt { E_s^2 -t_{\perp}^2/4 } ~~{\rm Exp }
(-{1\over N(0) U })
\ee
where $E_s$ is some cut off energy below which the system undergoes
SDW transition. Equation ( 18 ) clearly shows that the SDW gap
reduces with the interlayer tunneling matrix element. The SDW
gap at zero temperature without interlayer tunneling can be
reproduced from equation (18) by setting $t_{\perp}=0$. For a
small but finite value of $t_{\perp}$, $
G(k)^{t_{\perp}\ne 0} /G(k)^{t_{\perp}}=0$ is plotted (Fig.1) for different
values of $k_{x}$ and $k_{y}$. It is shown that the SDW gap
reduces upto $98 \%$ of its original value (in absence of
$t_{\perp}$) in certain directions of the FS and thereby the SDW
gap becomes anisotropic.
The parameter values chosen are, $E_s = 0.3 eV$, $N(0)V=0.5$ and
$t_{\perp}$ varies from zero to $0.1 eV$. With the above
parameter values we get a SDW transition temperature ( mean field )
of $400^o K$.  The SDW order parameter $ G$ signifies the net amount of
magnetic moment
(in the z-direction for longitudinal SDW) at each site.  Therefore,
the interlayer tunneling have similar effect as magnetic impurity in ordinary
superconductors, i.e the interlayer tunneling will cause SDW
pair breaking leading
to the suppression of SDW order parameter. The suppression of the inplane
SDW ground state in turn enhances the superconducting correlations,
simply because the more the number of SDW pair breaks the more number of
quasiparticles becomes available for superconcting pairing, which we
shall discuss in the next section.

 The k-dependence of the SDW gap
essentially depends on the anisotropic nature of the
$t_{\perp}(k)$. It is known from electronic structure
calculations \cite{mat} of high-$T_c$ materials  (each $Cu-O$
layers being two-dimensional) that the two layers touch along the
$\Gamma M$ line. The point $M$ corresponds to $(\pi/a,~ \pi/a)$.
The largest splitting of the hybridized bands of the two layers
is seen to be at the point $X$, which is $(\pi/a,~0)$. Following
the above symmetry, Chakravarty \cite{che} et al, chosen the form
of $t_{\perp}(k)$ as
\be
t_{\perp}(k)= {t_{\perp} \over 4}[\cos(k_x a) - \cos(k_y a)]^2
\ee
We use the same form of $t_{\perp}(k)$. Now, combining (18) and
(19) it is easy to see that the SDW gap will be maximum along
the $k_x = k_y$ line whereas it will be minimum along the
directions where $t_{\perp}(k)$ is maximum (i.e, the points $(0,
\pm \pi/a), ~ (\pm \pi/a, 0)$) (cf. Fig.2). Therefore, the SDW gap will
become smaller and smaller in certain regions of the FS, whereas
it will retain its unperturbed magnitude in certain
other directions.
Further as we have already mentioned in sec.2. that the
SC-pairing will take place predominantly between the SDW quasi
particles in the gap free regions of the FS. So the SC gap peaks up in the
regions where the SDW gap gets suppressed and vice-versa.
\section{Superconducting State}

	The effective interaction between the SDW quasi
particles and the fluctuations of the collective modes of the
SDW state will give rise to a new kind of electron-amplitudon (phason)
interaction This  will be responsible for superconductivity under
the present mechanism. Such an effective interaction considering
the phase and amplitude fluctuations of the order parameter had
already been constructed by Behera et al, \cite{6} using linear
response theory
 of density fluctuations.
In the present picture similarly, the
interaction Hamiltonian due to the amplitude
and phase fluctuations of the SDW order parameter in the $H_0$
part of the Hamiltonian in (2) is developed. However, the
SDW order parameter
$G$ in eqn(2) has to be replaced by the self-consistently
calculated $G$ in eqn(18), such that the essential effect of
the interlayer tunneling is
 taken in to account in intermediate SDW state.
	The interaction Hamiltonian
can be written as [see ref.6],
\be
H_I=U\sum_{k,q,\sigma}\sum_{i=1}^{2}
\hat\Psi_{k+q,\sigma}^\dagger \hat \sigma_i
\hat\Psi_{k,\sigma} (d_q^{i} +d_{-q}^{{i}^\dagger})
\ee
where  $d_i$ is the annihilation operator
for the amplitude and phase
fluctuation modes of the SDW state. The above interaction Hamiltonian
can be reduced to an attractive effective electron-electron
interaction by usual second order perturbation theory. As is
clear from eqn(20)
that the form of the effective interaction
between the SDW quasi particles
will be quartic in $\Psi_{k,\sigma}$. In deducing the effective
interaction
(second order process) between the SDW quasi particles from
both (the upper
and the lower SDW) bands the intraband interaction terms
like $\gamma^{v^\dagger}
\gamma^{v^\dagger}\gamma^v \gamma^v$ and
$\gamma^{c^\dagger}\gamma^{c^\dagger}\gamma^c \gamma^c$ are neglected. The
details of the mean field theory leading to pairing Hamiltonian
in case of charge density wave (CDW) superconductors (non-cuprate high
$T_c$ superconductors )
is discussed in \cite{hng}.
We obtain the mean field pairing Hamiltonian derived from the
effective electron-electron
interaction as,
\be
H_{eff}=\sum_{k,\sigma}\sum_{i=1}^2 E_k[\gamma_{i k \sigma}^{c^\dagger}
\gamma_{i k \sigma}^c + \gamma_{i k \sigma}^{v^\dagger}\gamma_{i k \sigma}^v ]
+\sum_k \sum_{i=1}^2 \Delta (k) [\gamma_{i -k\downarrow}^c
\gamma_{i k\uparrow}^c
+ \gamma_{i -k\downarrow}^v \gamma_{i k\uparrow}^v + h.c ]
\ee
where $\Delta (k)$ appears as the effective SC-gap in the electronic energy
spectrum as $$e_k= \sqrt{{E_k}^2 + {\Delta_{SC} (k)}^2}$$.

     The anisotropic SC order parameter $\Delta_{SC}(k)$ is obtained as,
\be
\Delta_{SC}(k)= -{1 \over N} \sum_{k^\prime} [\lambda \sin^2 (\phi_k +
\phi_{k^\prime}) + \lambda^\prime \cos^2 (\phi_k +\phi_{k^\prime})]
\langle \gamma_{i -k^{\prime}\downarrow}^c \gamma_{i k^{\prime}\uparrow}^c
\rangle
\ee

where $\lambda (\lambda^\prime)$ are the dimensionless effective
electron-electron coupling constants (proportional to $U^2$)
and is equal to $\lambda,\lambda^{\prime}=
{\Omega_{AM}\over (E_k\mp E_{k^{\prime}})^2-\Omega_{AM}^2}$,
where $\Omega_{AM}$ being the maximum frequency
of the SDW gap fluctuations (amplitudon) given by $\Omega_{AM}$ $= 2~G$.
The order
parameter $\Delta_{SC}$ is assumed to be real and invariant under the
transformation $v \leftrightarrow c$. Self-consistent evaluation
of the correlation
function in the above equation yields the SC-gap equation as an
integral equation given by,
\be
\Delta_{SC} (k) = \sum_{k^\prime} [\lambda_1 +\lambda_2 {(\epsilon_k - \mu)
(\epsilon_{k^\prime} - \mu) - G(k) G(k^\prime) \over {E_k
E_{k^\prime} }}] (\Delta_{SC} (k^\prime)/e_{k^\prime})
\tanh (\beta e_{k^\prime}/2)
\ee
where $\lambda_{1(2)}=\lambda\pm \lambda^{\prime}$. The gap equation
in general is very difficult to solve analytically. We solve the
 gap equation in the low doping limit, i.e $\epsilon_k -\mu
\rightarrow 0$, implying that the injected carriers will stick
to the band gap edges with very low mobility (see Schrieffer et al \cite{5}).
With this approximation, the gap equation becomes,
\be
\Delta_{sc}(k)=\sum_{k^\prime} [ (\lambda_1-\lambda_2)
+\lambda_2 {(\epsilon_k-\mu)(\epsilon_{k^\prime}-\mu)
\over {G(k)G(k^\prime)}} ] {\Delta{sc}(k)\over e_{k^\prime}}
{\rm tanh }({\beta e_{k^\prime}\over 2})
\ee
It is clear that, the $T_c$ expression would have been like usual BCS
if only the first term in the r.h.s. was present, but it
will vary with SDW gap parameter G(k) because of the second term.
We obtain,
\be
K_B T_c=1.14 \hbar \Omega_{AM} [ Exp(-{1\over (\lambda_1-\lambda_2)N(0)}
+{\lambda_2 \delta^2\over 4 G^2N(0)}) ]
\ee
where $\delta$ is the doping concentration, satisfies ${1\over N}
\sum_{k\sigma} \langle c_{k\sigma}^\dagger c_{k\sigma} \rangle =
1-\delta$ and fixes the chemical potential $\mu$.
 From equation (25) we see that, as $t_{\perp}$ increases, the SDW gap G
decreases, which in turn increases the superconducting transition temperature.
For larger $t_{\perp}$, the SDW gap becomes smaller and smaller,
and hence greater number of low energy
quasiparticles becomes available for pairing.
This leads to the increase of superconducting transition temperature with
increasing $t_{\perp}$ (cf. Fig. 3).
Therefore, maximum $T_c$ is obtained only for non zero $t_{\perp}$ in this
novel mechanism. This is contrary to the situation where, single particle
tunneling operates between two planar superconductors. There $T_c$
decreases with increase of $t_{\perp}$.

\section{Conclusions}
	We propose a physically realistic theoretical model
(called `modified spin bag model') for high temperature
superconductivity  --- an appropriate modification over
Schrieffer's spin bag model \cite{5}. Inclusion of single
particle interlayer tunneling in the intermediate metallic SDW
state incorporates  a more realistic  three dimensional
situation. We presented in detail an effective perturbative approach
for the transverse coupling and thereby the lower dimensional
fluctuations are treated more rigorously. We also provide a
quantitative description of the SC-pairing mechanism (modified
spin bag scenario).

In the stoichiometric compounds, the CuO layers are considered as SDW
insulators due to the presence of perfect nesting of Fermi surface
and intermediate strength coulomb repulsion.  In order to get SC one needs
to dope the system with free charge carriers over the SDW state.
The Fermi surface deviates from
perfect nesting with doping, and SDW gap vanishes in most part of the Fermi
surface, leaving small pockets of gapped regions where nesting is still
preserved. The collective modes of the surviving SDW gap ( amplitude
and phase modes) mediates the pairing interaction between the low energy
SDW quasiparticles available near the gapless regions. This leads to
superconductivity.
 We point out, that in the normal state with a pseudo gapped
Fermi surface, there are low energy quasiparticles available in the
gap free regions. These are almost like free charge carriers and are avilable
for tunneling across the planes. The single particle tunneling between the
layers,
essentially acts as a SDW pair breaking perturbation, leading to further
suppression of the SDW gap. Hence the number of low energy quasiparticles
available for superconducting pairing increases, leading to an
increase in the transition temperature. So the maximum $T_c$ is obtained
for a nonzero value of $t_{\perp}$.
The increase of $T_c$ with single particle tunneling amplitude,
is exactly opposite to what happens in the case of,
two planar superconductors coupled by a tunneling term.
There $T_c$ decrease with increase in $t_{\perp}$.

	As mentioned in Sec.2, there exists two distinct regions
of the FS in the metallic SDW state --- the gapless regions due
to loss of perfect nesting in certain directions and the gapped
region where nesting of the FS survives. The gapless regions of
the FS is nothing but the nested Fermi liquid \cite{ruv}. It is
possible to show that in the large doping limit most of the
regions of the FS is free electron like in the normal state and the
nature of the SDW quasi particles e.g, life time etc. in the
gapfree regions are nested Fermi liquid like. As a result
scattering of light by the charge carriers is very large in the
normal state. This will naturaly lead to a large constant
background intensity in the Raman spectrum. The calculated
electronic Raman scattering intensity shows suppression on going
from normal to SC state in accordance with the experimental
results and will be published separately elsewhere \cite{man}.

  Furthermore in our mechanism, since the SDW gap survives in
some parts of the Fermi
surface, even below  $T_c$, this would explain the origin of short
range antiferromagntic correlations observed in the cuprates below the $T_c$.
It is observed in high $T_c$ materials \cite{alt}, that several
phonon modes shows anomalous
softening and narrowing of line width, even above the $T_c$.
In NMR relaxation experiments it is observed to show a spin gap
like feature, leading to decrease of relaxation rate much
above the transition temperature. These can be identified with
the existence of
SDW gap in some parts of FS in our model.

\newpage
\begin{figure}[p]
FIGURE CAPTIONS

\caption{ The ratio of $G(t_{\perp})$ and $G(t_{\perp}=0)$
is plotted as a function of momenta $k_x$ and $k_y$. $G(t_{\perp})$
reduces to $98 \%$ of  $G(t_{\perp}=0)$ along $\Gamma X$ points.}

\caption{
The SDW and superconducting gaps in arbitrary units
are plotted versus $t_{\perp}$ along $\Gamma X$ points.
The SDW gap shows suppresion whereas the superconducting gap
increases with $t_{\perp}$.}

\caption{The superconducting transition temperature is plotted
against $t_{\perp}$. The $T_c$ increases with $t_{\perp}$.}

\end{figure}

\end{document}